# Experimental Verification of Van Vleck Nature of Long-Range Ferromagnetic Order in Vanadium-Doped Three-Dimensional Topological Insulator $Sb_2Te_3$

Mingda Li[1,2*], Cui-Zu Chang[2*], Lijun Wu[3], Jing Tao[3], Weiwei Zhao[6], Moses H. W. Chan[6], Jagadeesh S. Moodera[2,5], Ju Li[1,4], and Yimei Zhu[3*]

[1] Department of Nuclear Science and Engineering, Massachusetts Institute of Technology, Cambridge, MA 02139, USA
[2] Francis Bitter Magnet Lab, Massachusetts Institute of Technology, Cambridge, MA 02139, USA
[3] Condensed Matter Physics & Materials Science Department, Brookhaven National Laboratory, Upton, NY 11973, USA
[4] Department of Material Science and Engineering, Massachusetts Institute of Technology, Cambridge, MA 02139, USA
[5] Department of Physics, Massachusetts Institute of Technology, Cambridge, MA 02139, USA
[6] The Center for Nanoscale Science and Department of Physics, The Pennsylvania State University, University Park, PA 16802-6300, USA

We demonstrate by high resolution low temperature electron energy loss spectroscopy (EELS) measurements that the long range ferromagnetic (FM) order in vanadium (V)-doped topological insulator $Sb_2Te_3$ has the nature of van Vleck-type ferromagnetism. The positions and the relative amplitudes of two core-level peaks ($L_3$ and $L_2$) of the V EELS spectrum show unambiguous change when the sample is cooled from room temperature to $T$=10K. Magnetotransport and comparison of the measured and simulated EELS spectra confirm that these changes originate from onset of FM order. Crystal field analysis indicates that in V-doped $Sb_2Te_3$, partially filled core states contribute to the FM order. Since van Vleck magnetism is a result of summing over all states, this magnetization of core level verifies the van Vleck-type ferromagnetism in a direct manner.

The breaking of time-reversal symmetry (TRS) and opening of surface bandgap of topological insulator (TI) is an essential step towards observing other quantum states [1-3]. When the TI's chiral Dirac surface state is gapped, a number of promising novel phenomena could be realized, including quantum anomalous Hall effect (QAHE) [3-10] where spontaneous magnetization and spin-orbit coupling lead to a topologically nontrivial electronic structure, and topological magneto-electric effect results through coupling between electric field and spin texture, which can potentially lead to low-power electrically-controlled spintronic devices [11-15]. There are two generic approaches to break the TRS: by magnetic proximity effect or by conventional transition metal (TM) doping [1, 16]. Doping TM impurities(*i.e.* V, Cr, Mn) into TI can induce a perpendicular ferromagnetic (FM) anisotropy, providing a straightforward method to open up the bandgap of the TI's chiral surface state and tune the corresponding transport properties [7, 8, 17-21]. In diluted magnetic semiconductors (DMSs) doped with TM atoms, the induced FM order in general originates from itinerant charge carriers [22, 23], *i.e.* Ruderman-Kittel-Kasuya-Yosida (RKKY)-mediated FM order. However, in magnetically doped TI, the itinerant carriers may destroy the QAHE by providing additional conduction channels [24], resulting in a leakage current [12, 25], which severely hinders magnetic TI for device applications. Therefore, a carrier-free yet long range mechanism to induce FM order in magnetically doped TI is highly desirable for progress towards device applications.

On the other hand, in magnetically doped TI, the first-principle calculations predicted that the insulating magnetic ground state can indeed be obtained by a proper choice of TM dopants, through van Vleck-type ferromagnetism in the absence of itinerant carriers [5]. Recently, Chang *et al.* [7, 8] has reported experimental observation of QAHE in magnetic TI Cr- and V-doped $(Bi,Sb)_2Te_3$, where the insulating FM order [21] excludes the RKKY-type interaction and indicates the FM mechanism to be of van Vleck-type as first-principle calculations predicted. In such a system, the inverted band structure in TI leads to the large matrix element of valence band [5, 21, 24], dramatically increasing the contribution to spin susceptibility. Since the van Vleck-type susceptibility is directly related with 2$^{nd}$ order energy perturbation [26], one could understand the van Vleck-type mechanism qualitatively from 2$^{nd}$ order perturbation theory, as shown in eq. (1),

$$E_0^{(2)} = \sum_n {}' \frac{\left|\langle 0|\mu_B(\vec{L}+g\vec{S})\cdot H|n\rangle\right|^2}{E_0 - E_n} \quad (1)$$

Here, $\sum'$ denotes the summation over all partially filled states. In this sense, the matrix element $\left|\langle 0|\mu_B(\vec{L}+g\vec{S})\cdot H|n\rangle\right|$ when $n$ becomes valence band is

huge, and contributes significantly to the spin susceptibility. On the other hand, eq. (1) also tells us that there are additional contributions from other partially filled states, enabling us to adopt a more direct approach to experimentally prove the van Vleck-type ferromagnetism without the carrier dependence, where electronic states in addition to itinerant electrons will contribute to FM order.

In this *Letter*, we report the magnetization of partially filled Vanadium (V) $2p_{3/2}$ and $2p_{1/2}$ ($L_3$ and $L_2$) core states, using low-temperature high-resolution Electron Energy Loss Spectroscopy (EELS). The condition for partial filling of core states is achieved when a high-energy incidence of electrons on a sample in a transmission electron microscope (TEM) excites a core electron to unoccupied states leaving a core-hole behind, giving to the energy-loss spectrum. Analyzing the fine structure of the energy-loss spectrum provides not only the information on the unoccupied local density of states, but also angular momentum, spin and chemical nature of the element. We find that by comparing with room temperature (*RT*) spectrum, the Te $M_{4,5}$ edge at $T$=10K shows no shift, while the V $L_3$ and $L_2$ peaks show a red shift as large as 0.6eV. In addition, there is a clear drop of $L_3/L_2$ peak intensity ratio. EELS simulation with FEFF 9.6 [27-29] shows that such a shift is a signature of onset of FM, which is independently verified through magneto-transport results, which shows anomalous Hall effect below the Curie temperature $T_C$~70K.

High-quality V-doped $Sb_2Te_3$ films are grown by molecular beam epitaxy (MBE) under a base vacuum ~$5\times10^{-10}$ Torr, where thin film $Sb_2Te_3$ (111) was grown on top of etched Si (111) substrates with V-dopants coevaporated from an electron beam source during TI growth. Ultrathin cross sectional TI film samples are fabricated through focused ion beam and post processing for high-resolution TEM studies. The EELS measurements were carried out using the doubled aberration corrected JEM-ARM200CF TEM, equipped with a cold field-emission gun and the state-of-the-art duel energy-loss spectrometer (Quantum GIF). V-doped $Sb_2Te_3$ samples still maintain a very good layered structure (Fig.1a), due to the likely case that V dopants tend to substitute Sb sites instead of creating interstitials [7]. The selected area electron diffraction (SAD) pattern (Fig. 1b) along [0001] zone-axis direction also verifies negligible influence of V-dopants to the crystal structure, in that the V-doped $Sb_2Te_3$ has almost identical lattice constant compared with that of pure $Sb_2Te_3$.

Fig. 2 shows the main result of this *Letter* comparing with the EELS spectra in the high-loss region at *RT* and low temperature *T*=10K. A simultaneous collection of both low-loss and high-loss spectra allows for a high-accuracy positioning of zero-loss peak thus accurate energy scale calibration. We see clearly that the $M_{4,5}$ edge of the Te element does not shift with temperature, while there are obvious peak position shifts (>0.5eV) for both V $L_3$ and $L_2$ peaks, accompanied with a decrease in $L_3/L_2$ ratio.

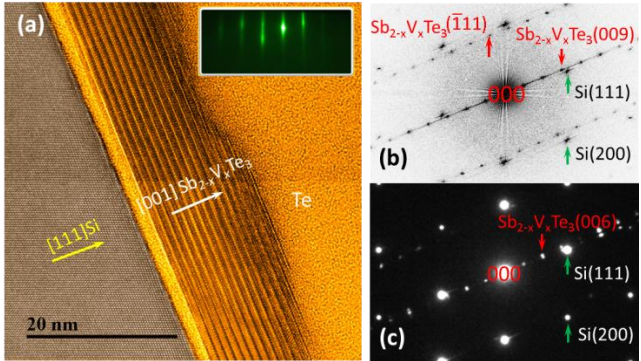

**FIG. 1** (color online). (a) High-resolution image of the V-doped $Sb_2Te_3$ sample S3 grown on etched Si substrate (bottom-left, brown region) viewed along [$\bar{1}\bar{1}0$] of $Sb_{2-x}V_xTe_3$. Another capping layer (top-right, yellow region) is mainly composed of amorphous Te protection layers. The upper left inset is a reflection high-energy electron diffraction (RHEED) image showing the ultrahigh crystalline quality of the MBE-grown film. (b) Diffractogram from (a). One set of spots as indicated by green arrows can be indexed as ($0\bar{1}1$)* pattern of Si, while the other set of spots indicated by red arrows can be indexed as ($\bar{1}\bar{1}0$)* pattern of a rhombohedral lattice with $a$=0.42 nm and $c$=3.03 nm which is basically the same as $Sb_2Te_3$ lattice, indicating negligible influence of V-dopants to the lattice. The [001] $Sb_2Te_3$ is slightly misaligned (~3°) with [111] Si (c) Select-area electron diffraction pattern from the area containing both film and substrate.

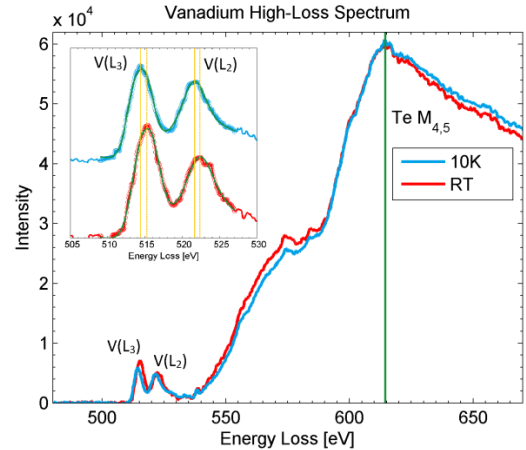

**FIG. 2** (color online). EELS spectra of V *L* and Te $M_{4,5}$ edges at *RT* (red curve) and 10 K (blue curve) for sample S2, normalized with Te $M_{4,5}$ edge intensity. The energy position of Te $M_{4,5}$ edge is invariant as temperature (green line), while there is a clear redshift of vanadium's $L_3$ and $L_2$ positions (yellow lines) and a drop of $L_3/L_2$ ratio. The energy scale has been accurately calibrated by simultaneously acquiring and aligning of the zero-loss peak.

In order to verify the observed peak shift, we measured three samples of $Sb_{2-x}V_xTe_3$ with different V concentrations and thicknesses, namely sample S1: 20 quintuple layer (QL) with $x$=0.08; S2: 20QL with $x$=0.16 and S3: 12 QL with $x$=0.08. The corresponding mean V-V distances in all the samples are thus $\geq 10$ nm. Since EELS is a spatially highly localized probe and there might be small non-uniformity of dopants, we collected 8 spectra at both 10K and $RT$ for each sample to reduce the measurement uncertainty. Furthermore, we use two different algorithms to extract the peak positions. The averaged V $L_3$ and $L_2$ peak positions and $L_3/L_2$ ratios for all the three samples are plotted in Figs. 3 (b-d).

All three samples show the same trend that the $L_3$ and $L_2$ peak positions at $T$=10K (blue and green dots in Figs. 3b and c) undergo a redshift compared with $RT$ (red and purple dots). For $L_2$ peaks, all three samples shift similar amounts ~0.4eV, while for $L_3$ peak positions, the shift ranges from 0.3eV (samples S1 and S3) to 0.7eV (sample S2). This indicates that at low temperature, certain mechanism which does not change Te states alters the $L_3$ and $L_2$ core states of V. The higher concentration of V tends to yield a stronger energy reduction of $L_3$ peaks. In addition, the $L_3/L_2$ ratio drops from ~1.4 to 1.1 (samples S2 and S3), indicating a possible change of electronic structure or even a phase transition [30].

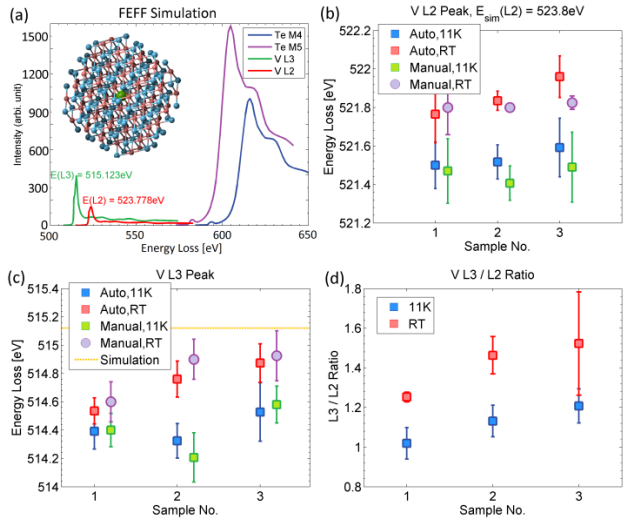

**FIG. 3** (color online). (a) FEFF simulation of the high-loss EELS spectrum of V-doped $Sb_2Te_3$, using a nanosphere (inset) with a scattering center in the middle. (b-d) Experimental EELS peak positions and shifts. (b) The V-$L_2$ peak positions, showing similar trend of redshift for all three samples. The two algorithms show consistent results (c) The V-$L_3$ peak positions, where sample S2 with highest V-concentration shows highest redshift. A horizontal yellow line marks the energy position from non-magnetic simulation, which is slightly higher than the $RT$ magnitudes. (d) The V's $L_3/L_2$ peak intensity ratio change. At $T$=10K, the ratio drops, which is also consistent with the simulation, where for non-magnetic system the ratio is even higher.

To understand the possible origin of the peak position shift and peak intensity ratio drop, we simulate the EELS spectrum with a non-magnetic V-doped TI nanosphere using FEFF9 (Fig. 3(a) inset). We take 1.0nm for full multiple scattering cutoff radius and 0.5nm for self-consistent-field cutoff radius to ensure convergence, with Hedin-Lundqvist self-energy and random phase approximation with core-hole correction. The resulting non-magnetic peak positions give $E(L_3)$ =515.1eV and $E(L_2)$ =523.8eV, which are both higher than the magnitude at $RT$. The higher V energy for a non-magnetic system is quite reasonable, since even at $RT$, there is already partial magnetization due to the field in the sample areas from the objective lens of the microscope. In other words, the redshift of V $L_3$ and $L_2$ peaks is consistent with a picture that non-magnetic system has even higher energy.

As shown in Fig. 3, the different amount of redshift for $L_3$ and $L_2$ edges is consistent with a temperature-independent Te $M_{4,5}$ position at ~615eV. Sample S2 with highest V concentration ($x$=0.16) shows the highest redshift, indicating that such redshift has an origin related to strong V-V interaction. Moreover, since the $L_3$ and $L_2$ has similar order of peak positions (around 515eV and 521eV), but different $l$-$s$ spin coupling configuration, the different redshift amount between $L_3$ and $L_2$ peaks further indicates that a spin-related process may play an important role in the V-V interaction.

Thus, the consistently observed trend at $RT$ and $T$=10K of the energy redshift of V's $L_3$ and $L_2$ edge and the decrease of $L_3/L_2$ ratio from non-magnetic simulations unambiguously indicates a change of electronic structure, while the very different redshift behavior between $L_3$ and $L_2$ peak together with a concentration dependence further indicating a magnetic origin from these core levels. Actually, this core-level magnetism for V-dopants could also be understood through crystal field theory. For an $l$=2 transition metal ion dopant (such as V), the character table of irreducible representation for full rotational group is shown in Table 1. The conjugacy classes are taken as the symmetry elements contained in TI's $D_{3d}$ group. Under TI's rhombohedral $D_{3d}$ crystal field, this irreducible representation $\Gamma_{full}^{(2)}$ becomes reducible, resulting in the lift of degeneracy and crystal field splitting.

| $\chi$ | E | $2C_3$ | $3C'_2$ | I | $2S_6$ | $3\sigma_d$ |
|---|---|---|---|---|---|---|
| $\Gamma_{full}^{(2)}$ | 5 | -1 | +1 | 5 | -1 | +1 |

**Table 1**. Character table of irreducible representation $\Gamma_{full}^{(2)}$ for the full rotational group. When crystal field is present, this becomes a reducible presentation and the degeneracy is lifted.

From Table 1 and character table of $D_{3d}$ group [31], we calculate the decomposition of the representation $\Gamma_{full}^{(2)}$ in $D_{3d}$ group as:

$$\Gamma_{full}^{(2)} = A_{1g} \oplus 2E_g \qquad (2)$$

i.e. instead of splitting to a 2-fold $E_g$ and a 3-fold $T_{2g}$ level which is the case of octahedral crystal field, an $l=2$ transition metal ion would split from a 5-fold level to two 2-fold levels ($E_g$ and $E'_g$) and one non-degenerate level $A_{1g}$ (Fig. 4) under TI's rhombohedral crystal field.

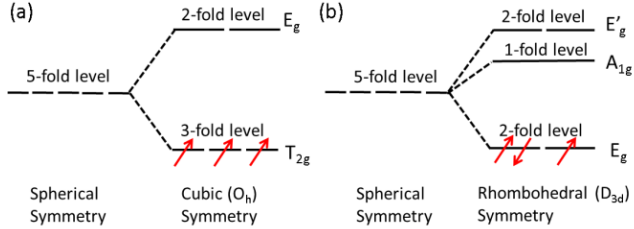

**FIG. 4** (color online). (a) Crystal field splitting in cubic crystal field. $T_{2g}$ levels allow the spin alignment from all three $3d$ electrons of V (red arrows), leading to a possible FM order. (b) Crystal field splitting under rhombohedral $D_{3d}$ crystal field. Since the energy only splits into a 2-fold level and only 1 electron has unpaired spin it is too weak to form a FM order.

Since V has electron configuration $[Ar]3d^34s^2$, this crystal field effect tends out to be important to explain why sole V's $3d$ electron states may not be sufficient to form FM order in TI. Unlike cubic crystal field where a 3-fold $T_{2g}$ state allows a parallel spin configuration (Fig. 4a), the 2-fold $E_g$ level and Pauli's exclusion principle only lead to single unpaired electron under rhombohedral crystal field. This becomes too weak to form a long range FM order by solely $3d$ valence states (Fig. 4b). Therefore, the FM order in V-doped TI may be mediated from other V-states, such as core states. This is fully consistent with our EELS results for $L_3$ and $L_2$ core states at $T=10$K.

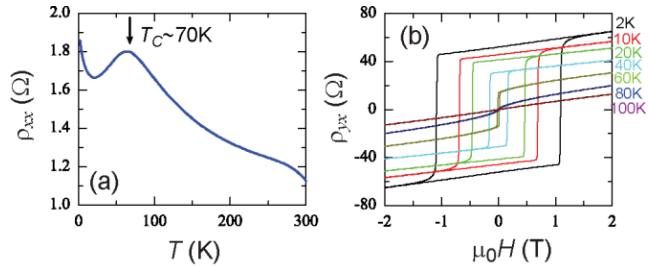

**FIG. 5** (color online). (a) Longitudinal resistance $\rho_{xx}$ as a function of temperature at zero-field. (b) Magnetic-field dependent Hall resistance $\rho_{yx}$. The hysteresis loop below ~70K is indicating an FM order.

In order to further demonstrate that the V-doped $Sb_2Te_3$ system is indeed FM at $T=10$K, we perform magnetotransport measurement for both longitudinal and transverse directions. Temperature-dependent longitudinal dc resistivity at zero magnetic field shows a resistivity hump at about 70K, indicating the onset of FM order (Fig. 5a). Figure 5b shows the hysteresis loop of the Hall resistance $\rho_{yx}$. The loop is closed above 70K, which is consistent with the $\rho_{xx}$ result. Hence we conclude that the Curie temperature $T_c$ to be ~70K, far above the EELS measurement temperature $T=10$K. This independently verifies the FM order of our V-doped $Sb_2Te_3$ sample.

In conclusion, we have demonstrated the van Vleck nature of FM order in V-doped TI $Sb_2Te_3$ using low-temperature high resolution EELS. An energy redshift is observed in V's $L_3$ and $L_2$ core states, which could be understood as a signature due to the onset of FM order, while the FM order itself is shown independently through magnetotransport measurement. The V-dopants' core-level contribution to the ferromagnetism in TI is thus in sharp contrast to the RKKY-type ferromagnetism, where only itinerant electrons contribute to the magnetic susceptibility regardless the core level states, but consistent with the picture of van Vleck-type ferromagnetism, where the susceptibility is a summation of contribution from all possible intermediate states. In this sense, although we could not exclude the contribution of RKKY interaction to the FM order from the band electrons, van Vleck mechanism, resulting from core levels and playing significant role in FM order, is observed unambiguously. Such a core-level contribution could also be understood from a crystal-field perspective, where three $3d$ electrons under rhombohedral crystal field could neither lead to FM order nor screen the contribution from the cores.

The work at the Brookhaven National Laboratory was supported by the U.S. Department of Energy, Office of Basic Energy Sciences under Contract No. DE-AC02-98CH10886. J.S.M. and C.Z.C would like to thank support from the STC Center for Integrated Quantum Materials under NSF grant DMR-1231319, NSF DMR grants 1207469 and ONR grant N00014-13-1-0301. Work at Penn State was supported by NSF MRSEC program under award numbers DMR-0820404 and DMR-1420620.

* Authors to whom correspondence should be addressed:
mingda@mit.edu (M. L.);
czchang@mit.edu (C. Z. C.);
zhu@bnl.gov (Y. Z.).